\documentclass[stslayout, noinfoline]{imsart}
\usepackage{geometry} 
\usepackage{graphicx}
\usepackage{amssymb, amsmath}
\usepackage{epstopdf}
\usepackage{natbib}
\usepackage[colorlinks,citecolor=blue,urlcolor=blue]{hyperref}

\graphicspath{{figures/}}

\begin{document}

\begin{frontmatter}

\title{The Convergence of Markov chain Monte Carlo Methods: \\
From the Metropolis method to Hamiltonian Monte Carlo}
\runtitle{The Convergence of Markov chain Monte Carlo Methods}

\begin{aug}
  \author{Michael Betancourt%
  \ead[label=e1]{betanalpha@gmail.com}}

  \runauthor{Betancourt}

  \address{Michael Betancourt is a research scientist in the Applied Statistics
           Center at Columbia University. \printead{e1}.}

\end{aug}

\maketitle


\end{frontmatter}

From its inception in the 1950s to the modern frontiers of 
applied statistics, Markov chain Monte Carlo has been one of 
the most ubiquitous and successful methods in statistical 
computing.  The development of the method in that time has 
been fueled by not only increasingly difficult problems but 
also novel techniques adopted from physics.  In this article 
I will review the history of Markov chain Monte Carlo from its 
inception with the Metropolis method to the contemporary 
state-of-the-art in Hamiltonian Monte Carlo.  Along the way 
I will focus on the evolving interplay between the statistical 
and physical perspectives of the method.

This particular conceptual emphasis, not to mention the 
brevity of the article, requires a necessarily incomplete 
treatment. A complementary, and entertaining, discussion of 
the method from the statistical perspective is given in 
\cite{RobertEtAl:2011}.  Similarly, a more thorough but still 
very readable review of the mathematics behind Markov chain 
Monte Carlo and its implementations is given in the excellent 
survey by \cite{Neal:1993}.

I will begin with a discussion of the mathematical relationship
between physical and statistical computation before reviewing
the historical introduction of Markov chain Monte Carlo and its
first implementations.  Then I will continue to the subsequent 
evolution of the method with increasing more sophisticated
implementations, ultimately leading to the advent of Hamiltonian 
Monte Carlo.

\section{From Physics to Statistics and Back Again}

At the dawn of the twentieth-century, physics became increasingly
focused on understanding the equilibrium behavior of thermodynamic 
systems, especially ensembles of particles.  For a system at 
constant temperature, $T$, that equilibrium behavior is completely 
characterized by the \emph{canonical} probability distribution,
\begin{equation*}
\pi (q, p) \propto \exp( - H(q, p) / k \, T ),
\end{equation*}
where $k$ is the Boltzmann constant, $q$ and $p$ are respectively 
the positions and momenta of the particles, and $H(q, p) = K(p) + U(q)$ 
is the \emph{Hamiltonian} defining the physical system. In particular, 
all physical observables of the equilibrium system become 
expectations with respect to this canonical distribution,
\begin{equation*}
\mathbb{E} [f] = \int \mathrm{d} q \, \mathrm{d} p
\, \pi (q, p) f(q, p).
\end{equation*}

Often physical observables are independent of the kinematics
of the system, reducing the calculation to expectations entirely 
on the configuration space,
\begin{equation*}
\mathbb{E} [f] 
= 
\int \mathrm{d} q \, \mathrm{d} p
\, \pi (q, p) f(q)
=
\int \mathrm{d} q \, \pi (q) f(q),
\end{equation*}
where the equilibrium distribution over configuration space
is given by
\begin{equation*}
\pi(q) \propto \exp(- U(q) / k \, T).
\end{equation*}
Ultimately equilibrated physical systems are equivalent to 
probabilistic systems, both specified by a given probability 
distribution with well-defined characteristics given by 
expectation values.

By exploiting this equivalence we can facilitate the
computation of physical observables.  The direct approach 
for computing observables of equilibrated physical systems, 
for example, is to simulate the dynamics long enough that
time averages over the dynamics converge to the equilibrium
observables.  Because of this equivalence, however, we 
can also employ techniques developed in statistics to avoid 
the costly simulations.

Moreover, the equivalence can be just as useful in
statistical problems.  Transforming a statistical system 
into a physical one allows us to utilize the corresponding
\emph{pseudodynamics} to improve existing computational 
methods.  The history of Markov chain Monte Carlo, and 
really much of modern statistical computation, is woven
from this interplay between physics and statistics.

\section{The Inception of Markov chain Monte Carlo}

A key insight from statistics is that every probability 
distribution admits a stochastic representation comprised
of a \emph{sampling procedure} that generates arbitrarily 
long sequences of points, 
$\left\{\theta_{1}, \ldots, \theta_{N} \right\}$, or
\emph{samples}, whose empirical average for any function, 
$f$, converges to the corresponding expectation value,
\begin{equation*}
\lim_{N \rightarrow \infty} 
\frac{1}{N} \sum_{n = 1}^{N} f (\theta_n)
= \mathbb{E} [ f ].
\end{equation*}
Without knowledge of the sampling procedure itself, the 
individual points in any such sample appear to be randomly 
distributed across parameter space, stochastically jumping 
between neighborhoods of high probability from one iteration 
to another.  

In particular, if we can generate samples from the canonical 
distribution, or its projection onto configuration space, 
then we can quickly calculate physical observables without 
having to resort to expensive simulations.  Unfortunately, 
as one considers increasing complex systems the generation 
of those samples becomes increasingly challenging.

\subsection{``...had to go to Monte Carlo''}

\emph{Exact} sampling procedures generate each point in a 
sample independently of the others.  Generating and publishing 
tables of exact samples from common probability distributions 
was a major effort of statistics in the early 20th century,
but the work was laborious and limited to very simple
distributions.  Still the method was a constant curiosity
for physicists.  Enrico Fermi, for example, would exploit 
the method to make what seemed like impossibly-quick 
predictions of experimental outcomes as early as the 1930s
\citep{Metropolis:1987}.  Similarly, Stanislaw Ulam would
play with sampling techniques to reason about the outcomes
of random games like poker and solitaire \citep{Eckhardt:1987}.

The computational technologies spurred by World War II
and the Manhattan Project, however, soon provided an
alternative means of generating samples from physical
systems.  Soon after the war ended, John von Neumann
realized the potential of the Electronic Numerical 
Integrator and Computer, or ENIAC, for automating the 
grueling task of generating random samples for estimating
physical observables.  

Together with Stan Ulam he developed computational methods 
for generating exact samples from arbitrary univariate 
distributions \citep{Eckhardt:1987} while his wife, Klari, 
worked with Nicholas Metropolis to program those algorithms 
into the ENIAC \citep{Metropolis:1987}.  At the same time 
Metropolis and Ulam formalized the initial mathematical 
foundations of the method \cite{MetropolisEtAl:1949}.
The program quickly became a success for the post-war
weapons program and the method was anointed ``Monte Carlo'' 
as an homage to the stories Ulam would tell about his uncle 
who was always asking for money to gamble at the infamous 
casino. 

Not much later the ENIAC was disassembled and moved
from the University of Pennsylania to Aberdeen Proving Ground
in Maryland, shortly delaying the work of von Neumann,
Metrolis, and Ulam.  In the meantime, however, Enrico
Fermi developed an entirely \emph{analog} version of
the method \citep{Metropolis:1987}.  A small trolley 
would trace the path of a neutron across a two-dimensional
material using tables of random numbers to determine
speed, direction, and collision times without the need
of expensive computer calculations.  The so-called
FERMIAC is still around, on display at the Bradbury
Science Museum in Los Alamos \citep{Bradbury:2017}.

\subsection{Markov Chain Reactions}

With von Neumann returning to Princeton after the war,
Metropolis continued on at Los Alamos, leading the group
applying Monte Carlo to study more and more advanced 
thermodynamic reactions.  As the reactions became more 
complex, however, generating exact samples became infeasible, 
even with the increasing computational power available 
from the upgraded ENIAC and its successor, MANIAC.

Along with Arianna Rosenbluth, Marshall Rosenbluth, Edward 
Teller, and Augusta Teller, Metropolis introduced a new 
scheme to generate \emph{correlated} samples from the
equilibrium distribution \citep{MetropolisEtAl:1953}. 
Arianna, an accomplished physicist herself, was responsible 
for the critical task of implementing the new algorithm on 
the MANIAC, while Marshall did the bulk of the methodological 
development \citep{Gubernatis:2005}.  The resulting scheme 
became known as the Metropolis method, although perhaps a 
more appropriate name would have been the Rosenbluth method.  

Regardless, the scheme introduced a stochastic dynamics that, 
while unrelated to the true dynamics of the physical system 
being studied, generates correlated samples from the 
equilibrium distribution on the configuration space.  The 
artificial pseudodynamics perturb the current state, $q$, to 
give a \emph{proposal}, $q'$, with is then accepted with 
probability
\begin{equation*}
a(q, q') = 
\min \! \left(1 , \frac{ \pi(q') }{ \pi(q) } \right).
\end{equation*}
If the proposal is rejected then the current state is 
repeated as a new point in the sample.

Provided that the nature of the perturbations does not 
depend on the current state, this procedure will generate 
a \emph{Markov chain} of points in the configuration space 
that defines a sample from the equilibrium distribution 
which can be used to estimate observables.  The generality 
of the approach suddenly made sampling methods feasible not 
only for the thermonuclear problems of immediate interest
but also for a variety of other challenging problems in 
physics and chemistry.

After decades of empirical success in the physical sciences, 
the Metropolis method was generalized by the statistician 
W. K. Hastings \citep{Hastings:1970} who realized that the 
proposal could be given by sampling from an arbitrary 
distribution, $Q(q \mid q')$, provided that the acceptance 
probability was modified to
\begin{equation*}
a(q, q') = 
\min \! \left(1 , 
\frac{ Q(q \,\, \mid q') \ \pi(q') }{ Q(q' \mid q \,\,) \ \pi(q) } \right).
\end{equation*}
The resulting \emph{Metropolis-Hastings method} is still 
to this date one of the most common methods of generating 
Markov chains for Monte Carlo estimation, a method now 
known as \emph{Markov chain Monte Carlo}.

The generality of the Metropolis-Hastings method and 
Markov chain Monte Carlo, however, does not guarantee 
reasonable practical performance.  In particular, if the 
Markov chain doesn't explore the target distribution well 
enough and fast enough then we will exhaust our computational 
resources long before generating a large enough sample to 
estimate expectations with any reasonable accuracy. For
the Metropolis-Hastings method this requires a proposal
distribution sufficiently well-suited to the target
distribution that each proposal strongly perturbs the
initial state without being rejected, ensuring large
distances between neighboring points in the subsequent
Markov chain.

\subsection{The Markov chain Monte Carlo Revolution in Statistics}

Despite Hastings' seminal contribution, Markov chain
Monte Carlo techniques were not yet strongly embraced 
by the statistics community.  Although exact sampling methods
were finding success in applied problems, for example in 
\cite{Rubin:1981} and \cite{DempsterEtAl:1983}, Markov 
chain Monte Carlo itself was largely considered untrustworthy 
as the correlations inherent in the samples initially made it 
difficult to build a rigorous theoretical understanding of when 
the method would yield well-behaved results.  That hesitancy 
to employ Markov chain Monte Carlo would soon change, however, 
with the introduction of a particularly useful implementation 
of the algorithm.

The rapid improvement and proliferation of computing in 
the 1970s stimulated not only physical scientists but also 
computer scientists, especially those working on the 
reconstruction of images corrupted by noise and deterioration.  
These reconstructions utilized statistical correlations 
between the black and white pixels in an image, correlations 
that looks suspiciously like those in spin lattices such as 
the Ising model that had been extensively studied in the 
physics literature.

Conveniently, methods for simulating from the equilibrium 
distribution of these lattices had already been developed in 
physics.  Chief amongst these was \emph{Glauber dynamics}
\citep{Glauber:1963}, which updates the orientation of each 
spin one at a time based on the configuration of its neighbors
on the lattice.

In 1984 Stuart and Donald Geman formalized this equivalency 
and applied Glauber dynamics to the problem of image 
reconstruction \citep{GemanEtAl:1984}.  The resulting 
\emph{Gibbs sampler} updated each pixel one at a time by 
sampling from the conditional distribution determined by 
the configuration of all of the other pixels in the image, 
generating a Markov chain of correlated pixel configurations 
distributed according to the desired distribution.

Not long after, Gelfand and Smith introduced the Gibbs
sampler to the statistics literature \citep{GelfandEtAl:1990}.
Having recently addressed the general problem of how to
decomposing a multivariate probability distribution into the 
univariate conditional probability distributions needed to 
implement a Gibbs sampler, the statistics field was particularly 
ripe for this implementation of Markov chain Monte Carlo.  Gibbs 
sampling became a rapid success, enabling a flurry of applied 
statistical analyses that had been up to that point infeasible.

The introduction of the software \emph{Bayesian inference Using 
Gibbs Sampling}, or BUGS, \citep{LunnEtAl:2000} was especially
critical to this revolution.  By allowing users to specify a 
large class of Bayesian posterior distributions bespoke to their 
analysis and then automating the application of Gibbs sampling, 
the software facilitated statistical analyses across a diversity 
of scientific fields.  Following its inception in 1989 and first 
public prototype in 1991, the software quickly became indispensable 
in many applied domains after its first stable release in 1995
\citep{LunnEtAl:2009}.

\section{Drifting Towards Scalable Markov Chain Monte Carlo}

The inertia of the Markov chain Monte Carlo revolution
in statistics carried over through the next few decades,
fueling the development of a rigorous understand of the 
performance and robustness of not only the Gibbs sampler 
but also Metropolis-Hastings samplers and Markov chain 
Monte Carlo in general \citep{MeynEtAl:2009}.  At the 
same time practical diagnostics such as the Gelman-Rubin 
statistic \citep{GelmanEtAl:1992} were introduced, 
testing the consistency of multiple Markov chains to 
identify the pathologies that had been identified as 
obstructions to accurate Markov chain Monte Carlo
estimation.  When carefully employed, these diagnostics 
could promote the robustness of Markov chain Monte Carlo 
in practical applications.

The success of these early Markov chain Monte Carlo 
implementations, however, soon contributed to their own 
demise.  Taking the success for granted, practitioners 
quickly advanced to increasingly complicated problems that 
saturated their capabilities.  

When targeting distributions with more than ten or so 
dimensions or non-trivial dependencies amongst the parameters,
Metropolis-Hastings implementations with simple proposals and 
Gibbs samplers become fragile or slow and quickly lose the 
ability to provide accurate estimators.  This diminishing 
performance motivated the development of the novel Markov 
chain Monte Carlo implementations needed to tackle the 
challenging problems to which practical interest had 
progressed.

\subsection{Langevin Monte Carlo}

The diminishing performance of the early Metropolis-Hastings 
implementations is not inherent to the Metropolis-Hastings method 
itself, but rather a consequence of the simple proposal distributions 
being used.  When applied to complex problems, simple proposals such 
as the random walk proposal first proposed in 
\cite{MetropolisEtAl:1953} either suffer from increasingly high 
rejection rates or increasingly small jumps \citep{RobertsEtAl:1997}.  
In either case we end up with a Markov chain that moves through 
parameter space so slowly that we exhaust our computational resources 
before we can adequately explore the target distribution.

Ultimately the problem is that the random walk proposal
generates an isotropic diffusion that in high-dimensions 
drifts away from the neighborhoods of parameters space
relevant to the target distribution.  If we let the diffusion 
drift for an extended time without applying a Metropolis-Hastings 
correction then we suffer from high rejection rates.  At 
the same time, however, if we apply the correction with high 
frequency then we interrupt the pseudodynamics and slow the 
exploration.  In other words, the pseudodynamics induced by 
random walk Metropolis proposals are incompatible with the 
structure of high-dimensional target distributions.

There are, however, more sophisticated diffusions that can
exploit the structure of a given target distribution. In 
particular, \emph{Langevin diffusions} utilize 
\emph{differential} information about the target distribution 
to confine the induced pseudodynamics to the relevant 
neighborhoods of parameter space.  The more directed 
exploration of this measure-preserving diffusion yields 
much more efficient exploration than the naive random walk 
of the original Metropolis algorithm.

One immediate issue with Langevin diffusions, however, is 
that they cannot be simulated exactly for most problems, 
and error in the numerical integration of the diffusion 
will bias the resulting samples.  \cite{RosskyEtAl:1978} 
noted that if numerical integration of the Langevin diffusion
is used as the proposal in a Metropolis-Hastings algorithm 
then the acceptance procedure will compensate for any 
numerical errors.  Provided that the gradients of the target 
probability density function are available, this approach
can drastically improve the performance of Markov chain
Monte Carlo.

\cite{RosskyEtAl:1978}, however, was largely unknown within 
the statistics literature, and Langevin methods were not
much considered until \cite{GrenanderEtAl:1994} was read 
before the Royal Statistical Society.  The paper introduced 
a discretized Langevin diffusion without any Metropolis-Hastings 
correction, which the authors argued was adequate for practical
problems.  Disagreement by Julian Besag in the formal discussion 
of the paper motivated Gareth Roberts and Richard Tweedie to 
undertake a theoretical analysis \citep{Roberts:2017} which 
demonstrated that the correction was critical to the stability 
of the algorithm \citep{RobertsEtAl:1996}.

This analysis confirmed that by carefully incorporating
differential information about the target distribution
into the proposal the \emph{Metropolis Adjusted Langevin 
Algorithm} drastically improved the performance and 
scalability of Markov chain Monte Carlo.  That said, the 
need to evaluate gradients of the target density proved
a substantial burden that limited the adoption of the 
algorithm both in statistics and science.

\subsection{Molecular Dynamics}

While Markov chain Monte Carlo was evolving, the continued
growth in computing technologies stimulated work on the 
direct simulation of physical systems \citep{FermiEtAl:1955, 
Dauxois:2008, AlderEtAl:1959, Rahman:1964}.

Symplectic integrators \citep{HairerEtAl:2006}, allowed larger 
and more complex physical systems to be simulated with enough 
accuracy that the resulting dynamical averages were not be too 
strongly biased from the equilibrium observables.  In challenging 
problems these \emph{molecular dynamics} methods typically yielded 
estimates much more quickly than competing Markov chain Monte Carlo 
approaches, although determining whether the bias induced by the 
numerical integration was indeed small enough for a given
application was a persistent challenge.

One particular complication with molecular dynamics, however,
is that because the physical dynamics are energy-preserving 
the dynamical averages yield only \emph{microcanonical} expectations 
conditioned on the given energy.  Quantifying the distribution 
over the energies themselves requires the difficult computation 
of the density of states.

\cite{Andersen:1980} introduced a workaround with the introduction 
of stochastic collisions into the physical dynamics.  The modified
dynamics would proceed largely as before, evolving along the initial 
energy level set for a random amount of time before a collision 
perturbed the momentum and hence the energy of the system.  For 
sufficiently strong perturbations the resulting dynamics explore the 
entirety of configuration space and in many cases ensure that the 
resulting dynamical averages converge to the equilibrium observables.

Interestingly, as the time between collisions becomes more frequent 
these dynamics converge to a Langevin diffusion.  The efficacy of
the modified dynamics, however, decreases in this limit as the
constant collisions limit how much the dynamics explore each energy
level  set.  The efficiency in which the unencumbered dynamics 
explores those level sets explains why molecular dynamics can often 
be more efficient than Langevin Monte Carlo methods.  In practice, 
the time between collisions has to be carefully tuned to ensure 
optimal performance.

\subsection{Hybrid Monte Carlo}

Although they evolved largely independently, researchers would
eventually identify that Markov chain Monte Carlo and molecular 
dynamics methods could be combined to yield an even more powerful 
technique.

Motivated by experimental verification of perturbative quantum 
chromodynamics in the 1970s, numerous groups entered the 1980s 
in a battle to tackle non-perturbative quantum chromodynamics.  
Most efforts considered \emph{lattice} methods that discretized 
the quantum fields on grids where the expectations over the space 
of field configurations became expectations over the discretized 
lattice configurations.  For bosonic fields the expectations over
the lattice configurations were amenable to existing Markov chain
Monte Carlo methods, but expectations over the complex-valued 
fermionic fields were still out of reach.

Initial work relied on \emph{quenching} approximations that
simply ignored the fermionic degrees of freedom in quantum 
chromodynamics.  By employing some of the most powerful 
supercomputers in the world, Gibbs samplers running across 
the discretized lattice were able to reasonably approximate 
the resulting bosonic expectations \cite{CreutzEtAl:1979}.  
In this early work the error induced by ignoring the fermionic 
fields was largely obscured by the error induced from the 
discretization of the fields themselves.  As computation 
improved and lattices became smaller, however, the fermionic 
fields would have to be taken into consideration.

The incorporation of fermionic fields was made possible when
\cite{WeingartenEtAl:1981} showed that expectations over 
fermionic fields could be calculated by augmenting a purely 
bosonic system with artificial momenta and taking expectations 
over the corresponding pseudodynamical system.  Unfortunately,
while expectations over these completely bosonic systems were 
amenable to Markov chain Monte Carlo the systems were too complex 
for existing samplers to be practical, even with the available 
supercomputing resources.

\cite{CallawayEtAl:1982} and \cite{PolonyiEtAl:1983} then
recognized that certain observables in quantum chromodynamics 
were equivalent to microcanonical expectations over the
pseudodynamical system introduced in \cite{WeingartenEtAl:1981}.  
Consequently these observables could be efficiently estimated 
with a molecular dynamics approach that simulated the 
pseudodynamics and took dynamical averages.  As with molecular 
dynamics, however, the numerical error induced by the numerical 
integration of the pseudodynamics was difficult to quantify and 
limited the utility of the computations.

\cite{BatrouniEtAl:1985} and \cite{Horowitz:1985} followed
with applications of Langevin diffusions to the pseudo-dynamical 
system, allowing for arbitrary observables to be estimated
and not just those equivalent to microcanonical expectations. 
\cite{ScalettarEtAl:1986} added a Metropolis correction
to the Langevin dynamics, independently proposing the Metropolis
Adjusted Langevin Algorithm yet again.

The stage was then set for the introduction of a \emph{hybrid}
method \citep{Duane:1985, DuaneEtAl:1986}.  Similar to
\cite{Andersen:1980}, this hybrid method simulated the 
pseudodynamics for an extended time before resampling the 
momenta and, consequently, the energy, allowing observables 
to be estimated with a single, prolonged time average. 

Finally, in order to correct for the inevitable error arising
from the numerical integration of the pseudodynamics, 
\cite{DuaneEtAl:1987} considered using the hybrid method as a 
proposal distribution for the Metropolis method.  As with the 
Metropolis Adjusted Langevin algorithm, the acceptance procedure 
rejected simulations that had accumulated too much error, exactly 
compensating for the inaccuracies of the numerical integrator. 
The resulting \emph{Hybrid Monte Carlo} algorithm then alternated 
between two steps -- a Metropolis transition driven by the 
pseudodynamics and a Gibbs transition that sampled new values 
of the auxiliary momenta parameters.

Before streamlining the name for publication, the authors had 
originally referred to this final approach as ``hybrid-guided'' 
Monte Carlo.  Behind the scenes, however, Pendleton and Roweth 
punned ``guided''  into ``guid'', the Scottish slang for ``good''
appropriate to their positions at the University of Edinburgh.
The method became know colloquially as ``guid'' Monte Carlo 
for its superior performance compared other algorithms 
\citep{Pendleton:2017}.

\section{Going with the Hamiltonian Flow}

Hybrid Monte Carlo quickly became the standard for lattice 
quantum chromodynamics calculations, but that would not be
its only application.  In an effort to read all papers every 
written about Markov chain Monte Carlo \citep{Neal:2017}, 
Radford Neal came across the method in the review 
\cite{Toussaint:1989} and recognized that the introduction 
of auxiliary momentum parameters could lift \emph{any} 
probabilistic system into a pseudodynamical system amenable 
to Hybrid Monte Carlo

After this realization Neal pioneered its application outside 
of physics, beginning with his seminal review of Markov chain 
Monte Carlo \citep{Neal:1993} and thesis on Bayesian neural 
networks \citep{Neal:1995}, and then culminating in a 
comprehensive review aimed at statisticians \citep{Neal:2011}.  

After his thesis Neal began to appreciate the importance of 
the pseudo-Hamiltonian system in the construction of the method
and transitioned to the revised name \emph{Hamiltonian Monte Carlo} 
\citep{Neal:2017}.  After many discussions about the method
with Neal, David MacKay adopted the new name in his influential
textbook \citep{MacKay:2003} which introduced the method to
many, including this humble author!  The improved naming was 
facilitated by Neal and MacKay not being aware that Hamiltonian 
Monte Carlo had already been used to denote unrelated algorithms 
in lattice quantum chromodynamics.

Eventually Hamiltonian Monte Carlo started to pique the interest 
of statisticians and practitioners.  In particular, the promise 
of a scalable method and the recognition that \emph{automatic 
differentiation} \citep{BuckerEtAl:2006} could automate the 
calculation of the necessary derivatives stimulated the development
of general-purpose software to facilitate the use of Hamiltonian 
Monte Carlo in practical applications.  

Initial attempts at software implementations, such as that 
in the Automatic Differentiation Model Builder, or \texttt{ADMB},
package popular in ecology \citep{FournierEtAl:2012}, were limited 
by the need to delicately tune the algorithm to achieve the 
promised performance.  It wasn't until 2011 when Hoffman and
Gelman introduced the No-U-Turn sampler capable of automatically 
tuning Hamiltonian Monte Carlo to achieve high performance in a
given problem \citep{HoffmanEtAl:2014} that the method really 
began to flourish in practice.

Along with a user-friendly probabilistic programming language 
\citep{CarpenterEtAl:2016} and a high-performance automatic 
differentiation package \citep{CarpenterEtAl:2015}, the 
No-U-Turn sampler formed the basis of \texttt{Stan}, a 
multi-environment platform for both specifying statistical 
models and estimating posterior expectations \citep{Stan:2017a} 
in the spirit of BUGS.  Today \texttt{Stan} has revolutionized 
applied statistics, admitting novel analyses in everything from 
political science, ecology, and medicine all the way back to 
astronomy and physics.  The dynamic Hamiltonian Monte Carlo 
methods pioneered in Stan have also been adopted in packages 
such as \texttt{ADMB} and \texttt{TMB} \citep{Monnahan:2017}, 
\texttt{PyMC3} \citep{SalvatierEtAl:2016}, and \texttt{NONMEM} 
\citep{Bauer:2017}, with implementations in numerous other 
packages currently in development.

Parallel to the exploitation of Hamiltonian Monte Carlo, 
theoretical analysis has illuminated the mathematical 
foundations of its success.  Ultimately the method is
driven by a \emph{measure-preserving flow} that rapidly
explores a given probability distribution, efficiently
aggregating the information needed to construct accurate 
Markov chain Monte Carlo estimators \citep{BetancourtEtAl:2016a}.  
The introduction of auxiliary momenta parameters is simply 
the unique means of lifting the given probabilistic system 
into a Hamiltonian system naturally equipped with the desired 
flow.  Exploiting this differential geometric understanding 
has illuminated both how to optimally implement the method 
in practice and how to diagnose potential biases.  A 
non-technical review of these developments is given in
\cite{Betancourt:2017a}.

Identifying measure-preserving flows over more intricate
spaces, such as tree spaces, in order to generalize 
Hamiltonian Monte Carlo is an active field of research.

\section{Conclusion}

From their inception, the development of Monte Carlo and 
Markov chain Monte Carlo methods has been constantly fueled 
by input from physics and related fields.  The equivalence 
between probabilistic systems and equilibrium physical systems 
provides a bridge that allows the fruitful exchange of ideas 
and techniques between physics and statistics.

This exchange extends beyond sampling methods.  Variational 
inference \citep{WainwrightEtAl:2008}, for example has proven 
successful for computation within certain classes of probability 
distributions.  Similarly, thermodynamic methods \citep{Cerny:1985, 
KirkpatrickEtAl:1983, MarinariEtAl:1992, Betancourt:2014} employ 
the structure of non-equilibrium systems to facilitate computation 
in multimodal problems.

Formalizing this connection and identifying unexploited dualities
promises to accelerate this exchange between the two fields and 
push both into new generations of computation.

\section{Acknowledgements}

I am extraordinarily thankful to Brian Pendleton, Simon Duane, 
Andrew Gelman, Tony Kennedy, Radford Neal, Christian Robert, 
Gareth Roberts, and Don Rubin for providing critical comments 
as well as personal recollections of the history presented in 
this article.

\bibliography{mcmc_history}
\bibliographystyle{imsart-nameyear}

\end{document}